\documentclass[usenatbib]{iaus}
\usepackage[pdftex]{graphicx}
\usepackage{subfigure}
\usepackage{amssymb}
\usepackage{amsmath}

\title[Internal wave breaking in solar-type stars] 
{Internal wave breaking and the fate of planets around solar-type stars}

\author[A. J. Barker \& G. I. Ogilvie]   
{Adrian J. Barker \and Gordon I. Ogilvie}

\affiliation{ Department of Applied Mathematics \& Theoretical Physics, \\ University of Cambridge, Cambridge, CB3 0WA, UK \\ email: {\tt ajb268@cam.ac.uk} }

\pubyear{2010}
\volume{271}  
\pagerange{000--000}
\jname{Astrophysical Dynamics: From Stars to Galaxies}
\editors{N. Brummel. A. S. Brun, M. S. Miesch, Y. Ponty, eds.}
\begin{document}

\maketitle

\vspace*{-0.4cm}

\begin{abstract}
Internal gravity waves are excited at the interface of convection and
radiation zones of a solar-type star, by the tidal forcing of a
short-period planet. The fate of these waves as they approach the centre of
the star depends on their amplitude. We discuss the results of numerical
simulations of these waves approaching the centre of a star, and the
resulting evolution of the spin of the central regions of the star and the
orbit of the planet. If the waves break, we find efficient tidal
dissipation, which is not present if the waves perfectly reflect from
the centre. This highlights an important amplitude dependence of the
(stellar) tidal quality factor $Q^{\prime}$, which has implications for the
survival of planets on short-period orbits around solar-type stars,
with radiative cores.
\vspace*{-0.2cm}
\vspace*{-0.3cm}
\end{abstract}
\vspace*{-0.2cm}
\firstsection 

\vspace*{-0.25cm}

\section{Introduction}
Tidal interactions are important in determining the fate of
short-period extrasolar planets and the spins of their host stars. The
extent of the spin-orbit evolution that results from tides depends on
the dissipative properties of the body. These are usually parametrized
by a dimensionless quality factor $Q^{\prime}$, which is an inverse
measure of the dissipation, and which in principle depends on tidal
frequency, the internal structure of the body, and the amplitude of
the tidal forcing. The mechanisms
of tidal dissipation that contribute to $Q^{\prime}$ in fluid bodies
are not well understood. We can generally decompose
the response to tidal forcing into an equilibrium tide, which is a
quasi-hydrostatic bulge, and a dynamical tide, which is a residual
wave-like response. Dynamical tides in radiation zones of solar-type
stars take the form of internal (inertia-) gravity waves (IGWs), which
have frequencies below the buoyancy frequency $N$. These have
previously been proposed to contribute to $Q^{\prime}$ for early-type
stars (e.g. \cite[Zahn, 1975]{Zahn75}). We consider a nonlinear
mechanism of tidal dissipation in solar-type stars, extending an idea
by \cite{GD98}. A short-period planet excites IGWs at the base of the
convection zone, where $N\sim 1/P$, where $P$ is its orbital period.
These waves propagate downwards into the radiation zone, until they
reach the centre of the star, where they are geometrically focused
and can become nonlinear. If their amplitudes are sufficient, convective
overturning occurs, and the wave breaks. This has consequences for the
tidal torque, and the stellar $Q^{\prime}$. We study this
mechanism, primarily using numerical simulations.

\vspace*{-0.6cm}

\section{Numerical results and their implications}

\begin{figure}[!ht]
 \vspace*{-0.65 cm}
\begin{center}
  \subfigure{
 \includegraphics[width=0.39\textwidth]{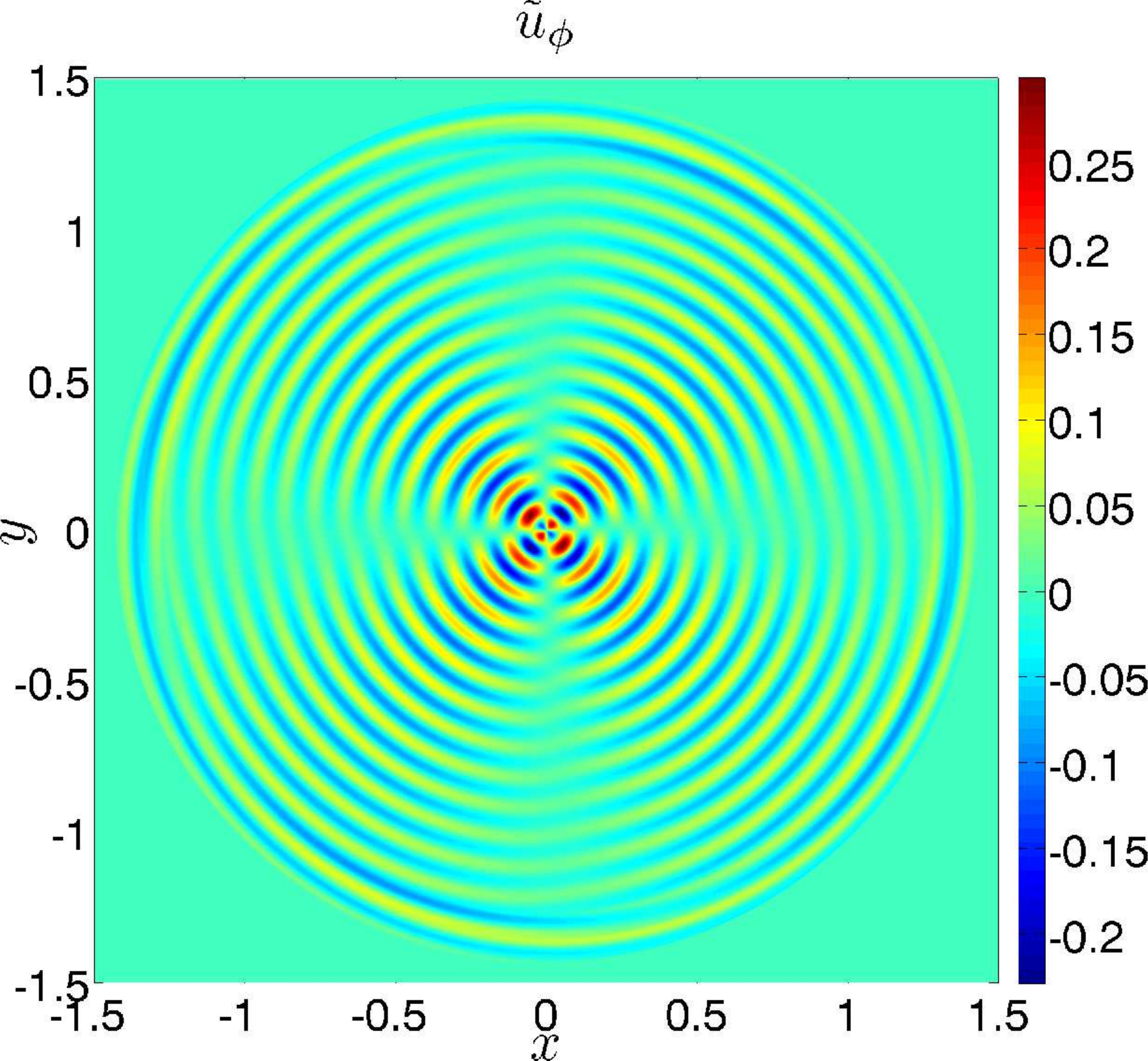} }
 \vspace*{-0.5 cm}
\subfigure{
  \includegraphics[width=0.39\textwidth]{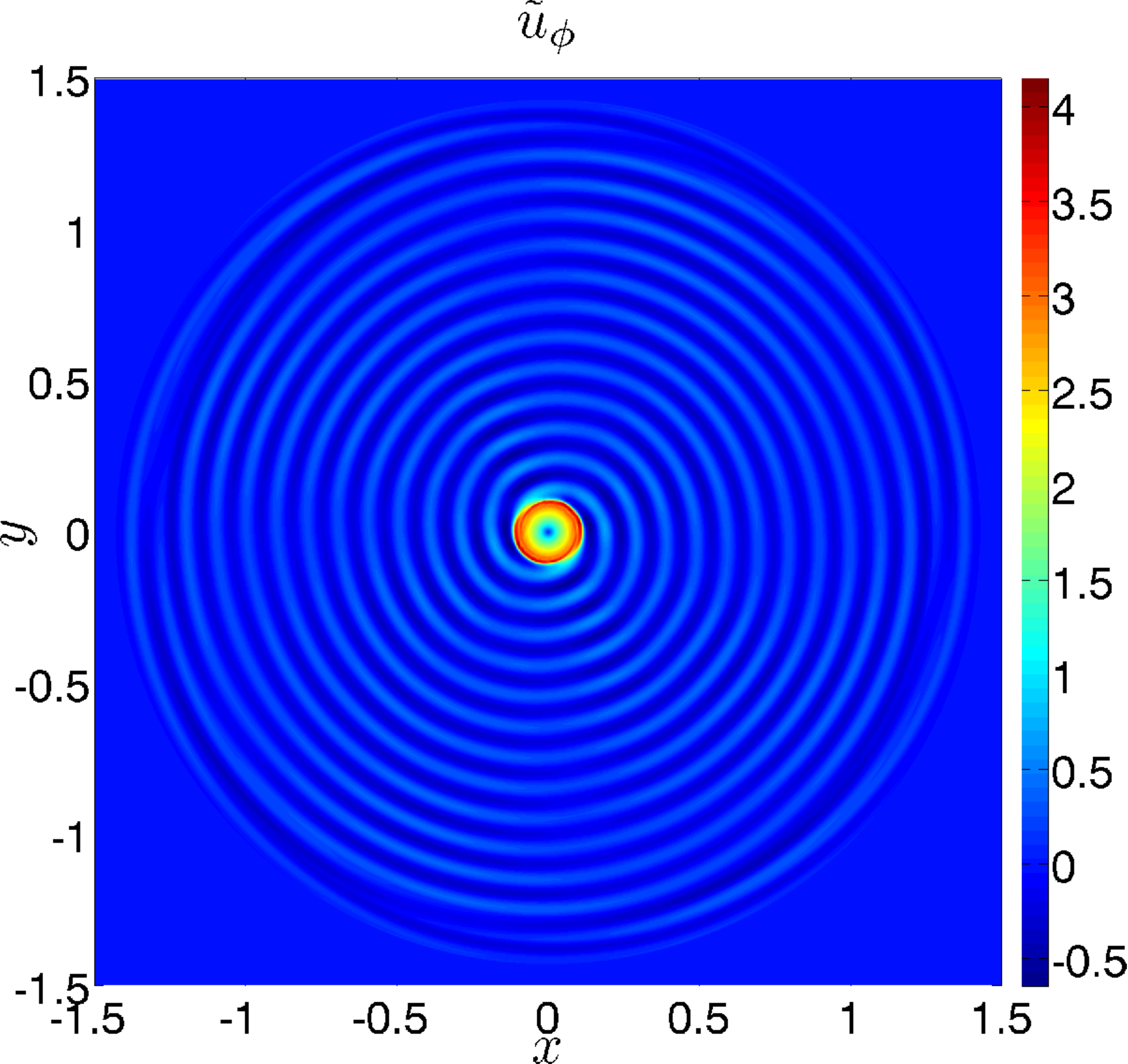} }
 \caption{Plot of the normalised azimuthal velocity near the
   centre of a solar-type star in two simulations with
   small (left) and large (right) amplitude forcing.}
   \label{fig1}
\end{center}
\vspace*{-0.25 cm}
\end{figure}

We solve the Boussinesq-type system of equations derived in
\cite{Barker2010}, which are valid in the central few per cent of
the radius of a solar-type star, where $g\propto r$ and $N= C r$. The
parameter $C$ measures the strength of the stable stratification at
the centre, and takes a value $C_{\odot} \approx 8.0
\times 10^{-11} \mathrm{m}^{-1}\mathrm{s}^{-1}$, for the current Sun. 
A Cartesian spectral code is used to solve these
equations, with our model being an initially non-rotating, 2D
cylindrically symmetric star\footnote{3D simulations with a
  spherically symmetric background have since been
  performed, which confirm these results.}. 
We artificially excite $m=2$ waves in the outer parts of the
computational domain, which are the dominant response for a planet 
on a circular, equatorial orbit. Fig.~\ref{fig1}
shows the azimuthal velocity, 
normalised to the farfield radial
phase speed of gravity waves in the simulation for a calculation with
small-amplitude (left) and large-amplitude (right) forcing.

If waves are excited with small-amplitude forcing, such that they
do not overturn the entropy stratification near the centre, then the waves
reflect perfectly from the centre of the star, and global modes
can form in the radiation zone. In this case efficient tidal
dissipation only occurs at discrete resonances \cite[(Terquem et al., 1998)]{T98}.

If the criterion \vspace*{-0.4 cm}
\begin{eqnarray}
\left(\frac{C}{C_{\odot}}\right)^{\frac{5}{2}}\left(\frac{m_{p}}{M_{J}}\right)\left(\frac{M_{\odot}}{m_{\star}}\right)\left(\frac{P}{1 
\;\mathrm{day}}\right)^{\frac{1}{6}} \gtrsim 3.3,
\end{eqnarray} \vspace*{-0.45 cm}

\noindent is satisfied, where $m_{\star,p}$ are the stellar and planetary
masses, waves are excited with large enough amplitudes so that
isentropes are overturned by fluid motions in the wave, within the
innermost wavelength. This leads to wave breaking and the deposition
of angular momentum, which spins up the mean flow to the orbital angular
velocity. This results in the formation of a critical layer, at which ingoing wave angular momentum is efficiently
absorbed, and the star is spun up from the inside out. This results in
efficient dissipation, leading to (for waves launched where $N^{2}$
increases linearly from the interface), \vspace*{-0.3 cm}
\begin{eqnarray}
  Q^{\prime}_{\star}  \approx 1.5 \times 10^{5} \left(\frac{P}{1 \;\mathrm{day}}\right)^{\frac{8}{3}},
\end{eqnarray} \vspace*{-0.45 cm}

\noindent which results in a rapid and accelerating planetary
inspiral, on a timescale on the order of Myr, for a
one-day Jupiter-mass planet around the current Sun.

This is a potentially important nonlinear mechanism of
tidal dissipation, which only operates in (solar-type) G
or K stars, with radiative cores. This process requires massive
planets or older/more centrally condensed stars. It is not in conflict with current observations of
extrasolar planets, and may explain the absence of massive close-in
planets around G-stars. This would not operate in F-type stars, with
convective cores, and so its absence may partly explain the survival of
massive close-in planets around F-stars, e.g. WASP-18.

\vspace*{-0.6cm}
\end{document}